# Is OpenAlex Suitable for Research Quality Evaluation and Which Citation Indicator is Best?

Mike Thelwall
Information School, University of Sheffield, UK. https://orcid.org/0000-0001-6065-205X
m.a.thelwall@sheffield.ac.uk
Xiaorui Jiang
Information School, University of Sheffield, UK. https://orcid.org/0000-0003-4255-5445

This article compares (1) citation analysis with OpenAlex and Scopus, testing their citation counts, document type/coverage and subject classifications and (2) three citation-based indicators: raw counts, (field and year) Normalised Citation Scores (NCS) and Normalised Log-transformed Citation Scores (NLCS). Methods (1&2): The indicators calculated from 28.6 million articles were compared through 8,704 correlations on two gold standards for 97,816 UK Research Excellence Framework (REF) 2021 articles. The primary gold standard is ChatGPT scores, and the secondary is the average REF2021 expert review score for the department submitting the article. Results: (1) OpenAlex provides better citation counts than Scopus and its inclusive document classification/scope does not seem to cause substantial field normalisation problems. The broadest OpenAlex classification scheme provides the best indicators. (2) Counterintuitively, raw citation counts are at least as good as nearly all field normalised indicators, and better for single years, and NCS is better than NLCS. (1&2) There are substantial field differences. Thus, (1) OpenAlex is suitable for citation analysis in most fields and (2) the major citation-based indicators seem to work counterintuitively compared to quality judgements. Field normalisation seems ineffective because more cited fields tend to produce higher quality work, affecting interdisciplinary research or within-field topic differences.

## Introduction

Citation-based indicators are widely used to support research evaluations of individuals, departments, universities and countries (De Bellis, 2009; Moed, 2006). This article investigates two separate issues with the same data: whether OpenAlex is a suitable database for citation analysis, and which is the best citation-based indicator.

### OpenAlex

Whilst citation analysts in the Global North have typically had to purchase licences to access the Scopus, Dimensions.ai, or Web of Science citation indexes, OpenAlex offers the same type of data for free and aims to provide greater coverage for the Global South (Priem et al., 2022). At the time of writing (February 2025) it also provided substantially greater coverage of books. In theory, this greater coverage should give better citation analysis results, because more data allows the results to be finer grained. Nevertheless, other factors are important, including document type classification (to ensure that only similar documents are compared), subject classifications (to ensure that documents are benchmarked only against others from the same subject), and citation extraction accuracy and comprehensiveness (because this is a non-trivial software engineering task). Thus, OpenAlex's greater coverage does not necessarily equate to greater utility for research evaluation tasks: this needs to be checked.

Based on previous research OpenAlex's document coverage is largely an expansion from Scopus, especially increasing the number of open access journals (Maddi et al., 2024) and data journals (Jiao et al., 2023) but with problems of metadata accuracy, including many articles without any indexed references (Alperin et al., 2024). OpenAlex's coverage of references is broadly comparable to Scopus and the Web of Science for documents that they all index, however (Culbert et al., 2024) so its expanded coverage is partly due to indexing documents without references. A particular problem with OpenAlex seems to be its classification of editorial contributions as standard journal articles (Haupka et al., 2024), which influences field normalized citation calculations (see below). It can also struggle to accurately classify the language of journal articles but seems to have wider coverage of non-English sources than Scopus (Céspedes et al., 2025).

*Citation-based indicators*

In parallel with the above OpenAlex motivation, a wide range of citation-based indicators have been proposed to help assess the impact or quality of journal articles. Whilst raw citation counts are widely available, it is unfair to compare citation counts between articles of different ages, because older articles have had longer to be cited. It is also unfair to compare citation counts between fields because some fields naturally tend to be more cited. Citation rate differences between two fields can occur because the number of references per article differs, or because one field tends to cite newer articles or non-article sources (van Raan, 2004; Glänzel et al., 2009; Moed, 2006). Thus, it is common in scientometrics to only compare citation counts (or citation rates for journals) within the same field and year. When citations need to be compared between fields and years, then the citation counts are typically converted into a field and year normalised rate or version. The most common is the Normalised Citation Score (NCS) (Waltman et al., 2011) or variants. For an article, this is its citation count divided by the mean citation count of all articles from its field and year. It has been argued that this is still unfair because the denominator (the field average citation count) is unduly influenced by the few highly cited articles (e.g., Brzezinski, 2015) and therefore all citation counts should be transformed to remove the skewing with $\ln(1+c)$ before any calculations are performed (Thelwall, 2017). This gives the Normalised Log-transformed Citation Score (NLCS). Percentile ranks are an alternative method to NLCS for dealing with skewed citation counts (Bornmann & Leydesdorff, 2013).

Despite the ostensible logic behind field normalisation, and many descriptive studies introducing or comparing indicators (e.g., Lundberg, 2007; Purkayastha et al., 2019; Waltman et al., 2012) or evaluating individual indicators against expert scores (Thelwall et al., 2023a), almost no study has systematically compared different indicators against a gold standard to demonstrate which is best or whether field normalisation is necessary. One partial exception is a small-scale comparison of indicators for 125 cell biology or immunology papers, where the gold standard was post-publication expert scores from F1000 website. Raw citation counts had the second highest corelation with expert scores (behind percentile rank in the subject area) (Bornmann & Leydesdorff, 2013). A larger scale analysis investigated 50,000 biomedical papers from F1000Prime (a newer name for F1000; subsequent names include Faculty Opinions and H1 Connect), finding that correlations between expert scores and a range of indicators were similar, but citation counts were not included so the value of field normalisation was not tested (Bornmann & Marx, 2015). A third partial exception is a study that shows that average citation counts correlated more strongly with a UK department's proportion of "world leading" research than did selected field normalised indicators for

Psychiatry, Clinical Psychology and Neurology (Fazel & Wolf, 2017). Nevertheless, the value of field normalization is unproven, and the relative merits of its component parts (e.g., the optimal granularity of the subject classification scheme used) are unknown.

From a wider perspective, citation-based indicators are typically used as either research quality or impact indicators, with the former seeming to be more accepted. The most common dimensions of research quality, at least as defined for peer or expert review, are originality, rigour, and significance (societal and scholarly) (Langfeldt et al., 2020). These three dimensions are all subjective in the sense that experts might score an article differently with a high degree of confidence and plausible justifications. Thus, the common practical solution is to accept the average or consensus of two or more peers or experts (e.g., Milyaeva & Neyland, 2023).

Subject classification schemes for science are essential for the field normalisation of indicators and a more accurate subject classification can expect to support better field normalised indicators. A comparison of classification schemes using long document reference lists as a gold standard for specialisms found that citation-based document classification tended to provide better results than journal-based classifications (Klavans & Boyack, 2017). No previous papers seem to have compared classification schemes by the extent to which the results match article-level quality judgements, however, presumably because of the scarcity of such data on a large scale.

## *Research questions*

The first goal of this paper is to assess whether OpenAlex is a credible platform for research evaluation indicators by comparing indicators derived from it with indicators derived from Scopus, an established and widely used citation index. The most likely sources of differences are (a) coverage of articles, (b) citation extraction accuracy/comprehensiveness, (c) subject/ classificaitons, and (d) document type classification. The second goal is to compare three alternative citation indicator formulae to identify the best: simple citation counts; NCS; and NLCS. It is not possible to test all these independently, but this paper will conduct a range of evaluations and then discuss alternative explanations for the results. The following research questions underpin this process.

- RQ1: Are OpenAlex citation counts better for research quality indicators than Scopus citation counts? The apparently greater coverage of OpenAlex may make its citation counts more useful, although if it covers low quality articles then their citations may degrade the overall value of citation counts.
- RQ2: Which is the best research quality formula: counts, NCS or NLCS?
- RQ3: Which is the best subject classification scheme: Scopus broad fields (overlapping), OpenAlex domains, OpenAlex fields, OpenAlex subfields, or OpenAlex topics?
- RQ4: Does OpenAlex or Scopus have the best classification of the "journal article" type for field normalisation formulae? OpenAlex has a less fine-grained document type classification, including editorials alongside full journal articles. It seems likely that this would compromise the value of indicators calculated from it.
- RQ5: Which is the best overall approach for calculating research quality indicators from OpenAlex and Scopus?
- RQ6: Are there field differences in the answer to RQ5?

# Methods

The research design was as follows.
1. Extract all journal articles from Scopus and OpenAlex for the years 2014-2020. These years were chosen to match the gold standards (see below).
2. Match the articles in Scopus and OpenAlex by DOI to create three sets: articles in Scopus; articles in OpenAlex; and articles in both (i.e., the intersection of the first two). Comparing results on the overlapping set will allow questions that do not involve coverage issues to be directly addressed (RQ1, RQ2, RQ3).
3. Calculate NLCS and NCS for all articles with all five classification schemes and on all three sets. This gives 5x3=15 indicators plus raw citation counts for OpenAlex and Scopus, giving a total of 32 indicators per article.
4. Calculate an article quality gold standard with ChatGPT for REF2021 articles. A secondary gold standard was calculated from departmental REF2021 average quality scores.
5. Correlate the 32 indicators against ChatGPT quality scores by UoA (Unit of Assessment – see below for an explanation) to identify the best overall approach and the influence of the component factors.

We had originally planned to use departmental mean REF2021 scores as the gold standard but, after analysing the results noticed that the gold standard was age-biased in favour of older articles. This occurred because higher scoring departments tended to submit a higher proportion of older articles to the REF. This gave an advantage to indicators that were not field normalised. Thus, individual article ChatGPT scores were used as an alternative. This has only a small age bias (Thelwall & Kurt, 2024) and is also preferable because it is direct, although using non-human "judgement" is a disadvantage.

## Data

The latest OpenAlex snapshot was downloaded on 1 December 2024. From this, documents of type "article" and with cross-ref type "journal-article" (as recorded in OpenAlex) were extracted along with their publication year, citation count, primary topic, field, subfield, and domain.

The metadata for all Scopus documents of type "journal article" and published between 2014 and 2020 was downloaded between 2 and 16 December 2024 to be approximately contemporary. The difference of up to 15 days results in a minor advantage for Scopus given that the most recent article analysed was from 31 December 2020 (a maximum 1% time difference, favouring Scopus, which does not affect the conclusions since OpenAlex tended to perform better). The data included publication year, DOI, citation count and All Science Journal Classification (ASJC) codes for each article (at least one each, but frequently multiple; these are derived from the publishing journal).

A joint dataset, "Both" was created from the subset of articles with a DOI in Scopus that were also in OpenAlex with the same DOI. DOIs were converted to lowercase before this comparison. Articles without DOIs were discarded because they could not be reliably matched.

## Scopus classifications

The Scopus ASJC classification is based on the fields of the publishing journal, as decided by subject specialist teams at Elsevier. For the 26 non-general broad fields in Scopus, there are two multi-purpose fields, called "all" and "misc" and the remaining AJSC narrow fields within the broad field are specialist (e.g., 2734 is Pathology and Forensic Medicine within the broad field 27 Medicine).

The Scopus AJSC classifications allow multiple classes per article, with 2 and 3 being common. For the field normalised indicators using the Scopus scheme, articles were assigned to each of their Scopus classes at full value. Although fractional counting could be used, it seems likely that the simpler whole counting approach is employed in practice.

## *OpenAlex classifications*

In contrast to Scopus (at least as accessible via its API), OpenAlex classifies articles individually, not by journal, and assigns them a single "primary" category rather than multiple equal categories. The OpenAlex data includes DOIs, subject classifications, and citation counts. The subject classifications include domains, field, subfields and topics. The field is one of 27 in the All Science Journal Classification (ASJC) list and the subfield is one of 252 from the same list. The topic is one of 4516 generated by a clustering of the articles. All articles are automatically assigned to these classifications from information including the "title, abstract, source (journal) name, and citations" (OpenAlex, 2025). Articles can be given multiple classifications but only the primary classifications are used here.

- **Domains**. 4: Life Sciences; Social Sciences; Physical Sciences; Health Sciences.
- **Fields**. 36. For example: Agricultural and Biological Sciences; Engineering; Health Professions; Social Sciences.
- **Subfields**. 252. For example: General Agricultural and Biological Sciences; Classics; Automotive Engineering; Speech and Hearing.
- **Topics**. 4516. For example: Tectonic and Geochronological Evolution of Orogens; Land Tenure and Property Rights in Agriculture; Structure and Agency in Social Theory; Medicinal Plants and Their Bioactivities.

## *Gold standards*

When assessing the value of an indicator, it is useful to have a "gold standard" in terms of a set of outputs with "correct" scores to compare the indicator against. This section describes the two (pseudo) gold standards used.

### Secondary: Departmental average research quality scores

Since citation-based indicators are typically used, explicitly or implicitly, as research quality indicators, the ideal gold standard would be a large collection of journal articles with agreed quality ratings that have been constructed with minimal citation data input. Since no such collection exists, a proxy was constructed from available Research Excellence Framework (REF) 2021 data, forming the secondary gold standard. REF2021 is the latest edition of the UK's national research evaluation process, which, amongst other things, involves 1,120 experts scoring 185,594 research outputs for research quality on a four-point scale: 1* (nationally relevant), 2* (internationally relevant), 3* (internationally excellent), and 4* (world leading) (2021.ref.ac.uk). The evaluation is primarily based on subjective assessments of the rigour, originality and significance of the outputs, with citations playing a very minor supporting role in some fields (Wilsdon et al., 2015).

Although the individual quality scores are confidential and have been destroyed, the percentage of scores awarded to the outputs of each submitting department are public, as are the identities of the department's outputs (results2021.ref.ac.uk). The departmental mean score was taken as a proxy indicator of the quality of each of a department's articles (or the mean of the departmental means for co-authored articles submitted by multiple departments). This is an indirect indicator but, unless there are systematic departmental biases, higher correlations with these proxies would associate with higher correlations with the (unknown) individual article scores, so the proxy is appropriate. One source of bias is if a department is stronger or weaker in books (or another non-article output type), but we circumvented this limitation with private access to journal-only departmental REF averages.

To ensure accurate matching, only articles with DOIs in the REF database were retained, and those with a publication date outside the year range 2014-2020 were removed. The secondary gold standard therefore consisted of all journal articles from REF2021 officially published 2014-2020, with a DOI matching both Scopus and OpenAlex (for comparability), and the REF2021 departmental average quality scores for journal articles.

The REF UoAs were used as a convenient database-agnostic set of field/subject categories for the correlation tests. This allows field differences to be reported. The REF is split into 34 broadly field-based UoAs for assessment purposes and these form a classification scheme that is independent of both Scopus and OpenAlex.

### Primary: ChatGPT article research quality scores

ChatGPT research quality scores were obtained for REF2021 articles to form the primary research quality gold standard. Many previous studies have shown that research quality scores from ChatGPT correlate positively with expert REF2021 departmental research quality scores in all fields (Thelwall & Yaghi, 2024; Thelwall et al., 2024) and the same for small-scale tests of individual article-level quality scores (Thelwall, 2024, 2025). Thus, in the absence of expert scores for individual REF2021 articles, ChatGPT provides the best available source of article-level research quality scores.

As shown by previous studies, the optimal method to obtain a useful research quality score from ChatGPT for an article is to configure it with system instructions from the relevant REF2021 expert review panel instructions, then submit the title and abstract of the article rather than its full text (Thelwall, 2025). The REF2021 instructions define research quality in terms of rigour, originality, and significance, then request an evaluation accompanied by one of the following scores: 1* nationally relevant; 2* internationally relevant; 3* internationally excellent; 4* world leading. This procedure was followed for all REF2021 articles without short abstracts (excluding the 10% of articles with the shortest abstracts, which included many short format submissions that are not full journal articles), with a program (github.com/MikeThelwall/Webometric_Analyst) used to automatically extract the scores from the reports. Articles with short abstracts were excluded because these are more likely to be mistakes or shorter document types that are not fully comparable with full articles. As part of this, all articles without abstracts in Scopus were excluded.

The requests were submitted to ChatGPT 4o-mini (gpt-4o-mini-2024-07-18) using its Applications Programming Interface (API) batch mode, with a separate chat session for each article score request. Although submitting multiple articles within the same chat session would save money by uploading the system instructions less often, this reduces the accuracy of the results (Thelwall, 2024). Each article was submitted five times in five separate non-consecutive sessions and the mean of the scores used as ChatGPT's estimate. Although the median or mode would be statistically better estimates of central tendency, the mean is preferable because it encodes ChatGPT's uncertainty about a score. Given that it has a strong preference for 3*, this uncertainty is the main device that can be used to leverage more fine-grained information in the underlying model. For example, a mean score of 3.2* indicates a small degree of uncertainty that 3* could be too low whereas 2.8* indicates a small degree of uncertainty that 3* is too high. For UoAs in which ChatGPT rarely gives 4* scores, from a correlation perspective, a 3.2* might be equivalent to a 4* in UoAs where it more frequently gives a 4* score (i.e., it might be a similarly top ranked score).

*Citation rate formulae*

As summarised in the introduction, it is unfair to directly compare citation counts with quality scores because older articles have had more time to attract citations, and different fields naturally have different citation rates. The standard approach to address both issues is to divide citation counts by the average citation counts of all articles from the same field and year, giving a field and year normalised citation count, such as the Normalised Citation Score (NCS) (Waltman et al., 2011). This is fairer than citation counts, for example because each article has a score of 1 if it has attracted an exactly average (mean) number of citations for its field and year. An alternative to this is the Normalised Log-transformed Citation Score (NLCS), which is the same except that each citation count c is transformed to ln(1+c) before all calculations. This reduces skewing in the original data and reduces the chance that individual highly cited articles could have a substantial influence on the results by changing the denominator of the normalisation calculation (Thelwall, 2017).

These indicators require a field classification system because this affects the choice of articles for their denominators, so both NCS and NLCS values are likely to change, perhaps dramatically, if the classification system changes. For this article NCS and NLCS values were calculated from Scopus, and OpenAlex citation counts, with all four OpenAlex classification schemes, and the separate Scopus ASJC. This gives multiple potential indicators per article to correlate against the gold standards.

For OpenAlex, normalised citation indicators were calculated separately with the OpenAlex 2014-20 set, the Scopus 2014-20 set and the overlapping 2014-20 "Both" set. Comparing results from these sets should indicate the influence of the main degrees of variation between the databases.

*Correlations*

The above process produces a large set of articles with gold standard quality indictors and a range of citation rate indicators. A simple way to answer the research questions would be to correlate the gold standard against each of the citation rate indicators and interpret the results. This is not ideal, however, since it is known that citations have substantially different values as research quality indicators, depending on the field. For example, they are relatively strong in the health, life and physical sciences but are weak in some social sciences and most arts and humanities (Thelwall et al., 2023a).

Instead of correlating citation-rate indicators against the gold standard overall, separate correlations restricted to each REF2021 UoA give finer-grained results and potentially reveal field-specific patterns. For example, perhaps OpenAlex is best for the arts and humanities (e.g., because of its book coverage) and worst for the health sciences. Using the REF UoAs seems fine-grained enough to identify major field-related patterns, without potentially giving a subtle advantage to Scopus or OpenAlex by using one of their classification schemes. Table 1 reports the main dimensions of variation between databases and the experiments included to test them.

Table 1. Dimensions of variation between citation indexes and indicators and tests to compare them. In each case the overall test is correlation against the gold standard quality scores with the UoA classification scheme.

| Dimension | Variation | Experiments (all possible variations) |
|---|---|---|
| Count source | Quality of citation count data, including the volume and importance of citations and the accuracy of citation extraction and matching. | Citation-based indicators calculated from Scopus citation counts compared to the equivalent calculated from OpenAlex citation counts. |
| Formula | Citation-based indicator formula | Citation counts, NLCS, and NLCS compared. |
| Norm reference set | Database coverage and document type classification from the perspective of (the denominators of) field normalised indicators. | Field normalised citation-based indicators compared between the OpenAlex document set, the overlapping (Both) document set and the Scopus document set. |
| Field classification scheme | Field classification scheme (both categories and categorisation) from the perspective of (the denominators of) field normalised indicators. | Field normalised citation-based indicators compared between classification schemes used for their denominators. |

*Indicator naming convention*

Thirty-two indicators were calculated, using all possible combinations of count source, formula, norm reference set, and subject classification scheme from those discussed above (Table 1). For convenience, the indicators all have a four-part label using the same notation (except that the count indicators only need two parts), as follows. Here, OpenAlex is abbreviated to OAlex to shorten labels.

- Part 1: Count source (OAlex or Scopus). This is the origin of the raw citation count for the article. For example, an OAlex count means the citation count for the article recorded in OpenAlex.
- Part 2: Formula (NLCS, NCS, or simple count). This is the mathematical formula (if any) used to transform a citation count into a field and year normalised variant.
- Part 3: Norm reference set (OAlex, Scopus, or Both). This is the set of articles that the given article is norm referenced against for the denominator of the NLCS or NCS calculations). For example, if OAlex is selected then the NLCS and NCS denominators will consist only of citation counts from OpenAlex (for articles within the same field).
- Part 4: Subject classification scheme (OAlex subfields, OAlex topics, OAlex fields, OAlex domains, Scopus fields). This is the subject classification scheme used to decide which papers are in the same field as the given article for the denominator of the NLCS or NCS calculations. For example, if OAlex subfields is selected then the NLCS and NCS denominators will only contain articles from the same OpenAlex subfield as the article in the numerator.

To illustrate the full schema, "OAlex|NLCS|Both|OAlex fields" corresponds to citation counts from OpenAlex, transformed into a field and year normalised citation score using the NLCS

formula, where the dominator of the formula consists of articles indexed by both Scopus and OpenAlex (with the same DOI) and classified into the same OpenAlex field.

# Results

There were 30,123,865 articles with DOIs from Scopus and OpenAlex combined for the years 2014-2020. All articles in Scopus contained a subject classification but 1,472,420 from OpenAlex did not (e.g., the case report 10.1001/jamaophthalmol.2019.0027) so the effective sample size was 28,651,445 journal articles. OpenAlex contained over twice as many articles as Scopus (Table 2).

Table 2. Number of journal articles in Scopus and OpenAlex by publication year.

| Publication Year | Scopus | OpenAlex | Both Scopus and OpenAlex | Scopus but not OpenAlex | OpenAlex but not Scopus | Total |
|---|---|---|---|---|---|---|
| 2014 | 1631903 | 3248341 | 1252466 | 379437 | 1995875 | 3627778 |
| 2015 | 1723950 | 3401330 | 1640627 | 83323 | 1760703 | 3484653 |
| 2016 | 1794387 | 3640650 | 1722085 | 72302 | 1918565 | 3712952 |
| 2017 | 1867502 | 3871401 | 1794360 | 73142 | 2077041 | 3944543 |
| 2018 | 1995996 | 4158490 | 1917273 | 78723 | 2241217 | 4237213 |
| 2019 | 2219803 | 4468685 | 2120222 | 99581 | 2348463 | 4568266 |
| 2020 | 2449836 | 4950270 | 2324066 | 125770 | 2626204 | 5076040 |
| **Total** | **13683377** | **27739167** | **12771099** | **912278** | **14968068** | **28651445** |

The gold standard contained 97,816 REF2021 journal articles with non-short abstracts (Table 3). Some articles are in multiple UoAs so the total of the UoAs is greater than the number of articles in the combined set.

Table 3. Sample sizes for both gold standards.

| UoA | Name | Articles |
|---:|---|---:|
| 1 | Clinical Medicine | 9727 |
| 2 | Public Health, Health Services and Primary Care | 3766 |
| 3 | Allied Health Professions, Dentistry, Nursing and Pharmacy | 9088 |
| 4 | Psychology, Psychiatry and Neuroscience | 7578 |
| 5 | Biological Sciences | 6056 |
| 6 | Agriculture, Food and Veterinary Sciences | 3016 |
| 7 | Earth Systems and Environmental Sciences | 3565 |
| 8 | Chemistry | 2649 |
| 9 | Physics | 3695 |
| 10 | Mathematical Sciences | 3046 |
| 11 | Computer Science and Informatics | 4103 |
| 12 | Engineering | 14445 |
| 13 | Architecture, Built Environment and Planning | 2265 |
| 14 | Geography and Environmental Studies | 3075 |
| 15 | Archaeology | 467 |
| 16 | Economics and Econometrics | 819 |
| 17 | Business and Management Studies | 9835 |
| 18 | Law | 1437 |
| 19 | Politics and International Studies | 1939 |
| 20 | Social Work and Social Policy | 2662 |
| 21 | Sociology | 1226 |
| 22 | Anthropology and Development Studies | 783 |
| 23 | Education | 2733 |
| 24 | Sport and Exercise Sciences, Leisure and Tourism | 2619 |
| 25 | Area Studies | 420 |
| 26 | Modern Languages and Linguistics | 745 |
| 27 | English Language and Literature | 524 |
| 28 | History | 710 |
| 29 | Classics | 78 |
| 30 | Philosophy | 446 |
| 31 | Theology and Religious Studies | 164 |
| 32 | Art and Design: History, Practice and Theory | 886 |
| 33 | Music, Drama, Dance, Performing Arts, Film and Screen Studies | 445 |
| 34 | Communication, Cultural & Media Studies, Library & Information Management | 810 |
| All | All the above, excluding duplicates | 97816 |

*Comparison between indicators on average across UoAs*

There are substantial differences between indicators in the extent to which they correlate with ChatGPT scores, on average across UoAs, although all correlations are positive (Figure 1). The two sets of bars in this figure are based on (a) the correlations calculated separately for each UoA (n=34) and then averaged across UoAs and (b) the correlations calculated separately for each UoA and year (n=34x7=238) and then averaged. Since there are 32 indicators calculated separately, the underlying data for Figure 1 consists of 34x32+238x32=8,704 separate correlations between indicators and average ChatGPT scores, collectively covering 97,816 articles (some multiple times) for each indicator. This section describes the results, and the Discussion speculates about the underlying causes. To avoid overcomplicating the presentation of results, the research questions are explicitly addressed only in the Discussion.

The relative value of the indicators varies according to whether the correlations are calculated for all years together or separately by year. The main analysis focuses on the calculation for all years together, which corresponds to how they would be used in the REF.

For all years analysed together (blue bars in Figure 1), the main patterns can be identified by comparing indicators where all factors are the same except one. For the source of data for the counts (RQ1) (either overall or in the NLCS or NCS formulae), the source with the highest correlation with ChatGPT is as follows, where x represents the component of the calculation being compared and italic indicates orders that are out of sequence compared to most of the other results.

- Scopus: None
- OAlex: All (e.g., x|Count||, x|NLCS|Both|OAlex fields)

Comparing indicator formulae (RQ2) for the one with the highest correlation:

- Counts: None (but comparable to the best NCS)
- NLCS: None
- NCS: All (e.g., OAlex|x|Both|OAlex fields, Scopus|x|Scopus|Scopus fields)

Comparing categorisation schemes (RQ3) (italics emphasise unusual orderings):

- OAlex|NCS|Both|x: Domains>fields>subfields>topics>Scopus fields
- Scopus|NCS|Both|x: Domains>fields>subfields>topics>Scopus fields
- OAlex|NLCS|Both|x: Domains>fields>*topics>subfields*>Scopus fields
- OAlex|NCS|OAlex|x: Domains>fields>subfields>topics
- Scopus|NLCS|Both|x: Domains>fields>subfields>*Scopus fields>topics*
- OAlex|NLCS| OAlex |x: Domains>fields >*topics>subfields*
- Scopus|NCS|Scopus|x: Scopus fields
- Scopus|NLCS|Scopus|x: Scopus fields

Comparing document classification scopes (RQ4):

- OAlex|NCS|x|OAlex domains: Both>OAlex
- OAlex|NCS|x|OAlex fields: Both>OAlex
- OAlex|NCS|x|OAlex subfields: Both>OAlex
- OAlex|NCS|x|OAlex topics: Both>OAlex
- OAlex|NLCS|x|OAlex domains: Both>OAlex
- OAlex|NLCS|x|OAlex fields: Both>OAlex
- OAlex|NLCS|x|OAlex subfields: Both>OAlex
- OAlex|NLCS|x|OAlex topics: Both>OAlex
- Scopus|NCS|x|Scopus fields: Both>Scopus
- Scopus|NLCS|x|Scopus fields: Both>Scopus

Overall (RQ5), it is clear that from the perspective of the highest overall correlation with ChatGPT for all years together (blue bars in Figure 1), that OpenAlex citation counts are preferable to those from Scopus, the optimal formula is NCS (although raw counts are a close second), the OpenAlex domains classification scheme is optimal and only documents classified by both Scopus and OpenAlex as journal articles should be included for the best results.

When the correlations are calculated separately by year (orange bars in Figure 1), the pattern is similar, but the main change is that raw citation counts are substantially better than any indicator. Thus, the main advantage of field and year normalisation formulae is that they make it fairer to compare articles from different years and not that they make it fairer to compare articles published in different fields.

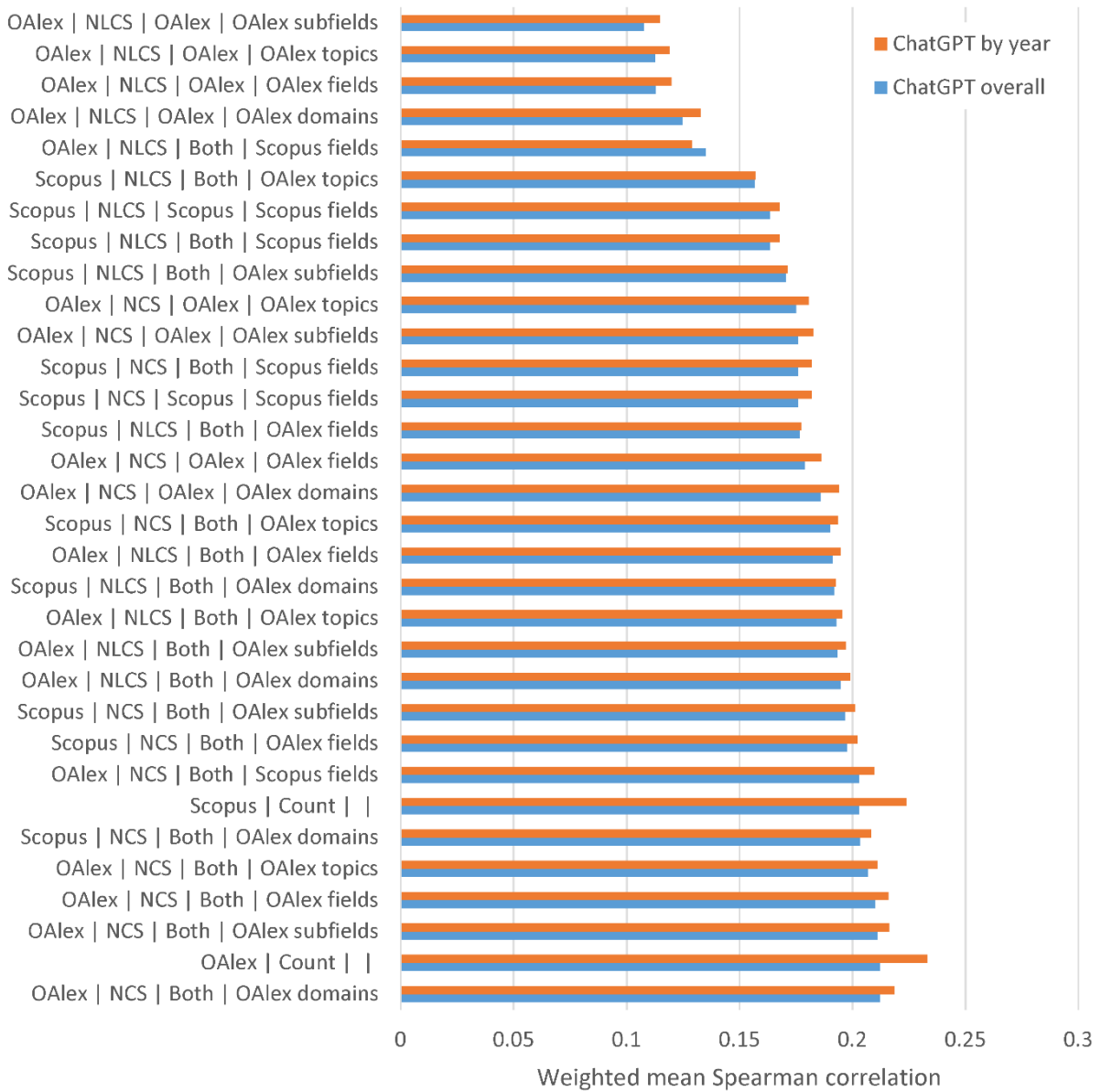

Figure 1. Weighted mean Spearman correlations between indicator values and ChatGPT scores for all qualifying REF articles, where the mean is taken over all 34 UoA correlations, weighted by the number of articles assessed in the UoA. The "overall" variant is based on all years combined within each UoA (i.e., one correlation per UoA), and the "by year" variant involves the correlations calculated separately for each UoA and year (i.e., one correlation per year x UoA), and averaged.

The order of the results changes little if average ChatGPT scores (i.e., the primary gold standard) are replaced with departmental average REF scores (i.e., the secondary gold standard) (Figure 2). Because higher scoring departments tended to submit a higher proportion of older articles and these tended to be more cited, there is a second order effect to favour citation counts for this comparison, which may be why the highest correlation is for raw counts rather than the field and year normalised NCS, and why calculating correlations by year makes less difference to the correlations than for the ChatGPT gold standard.

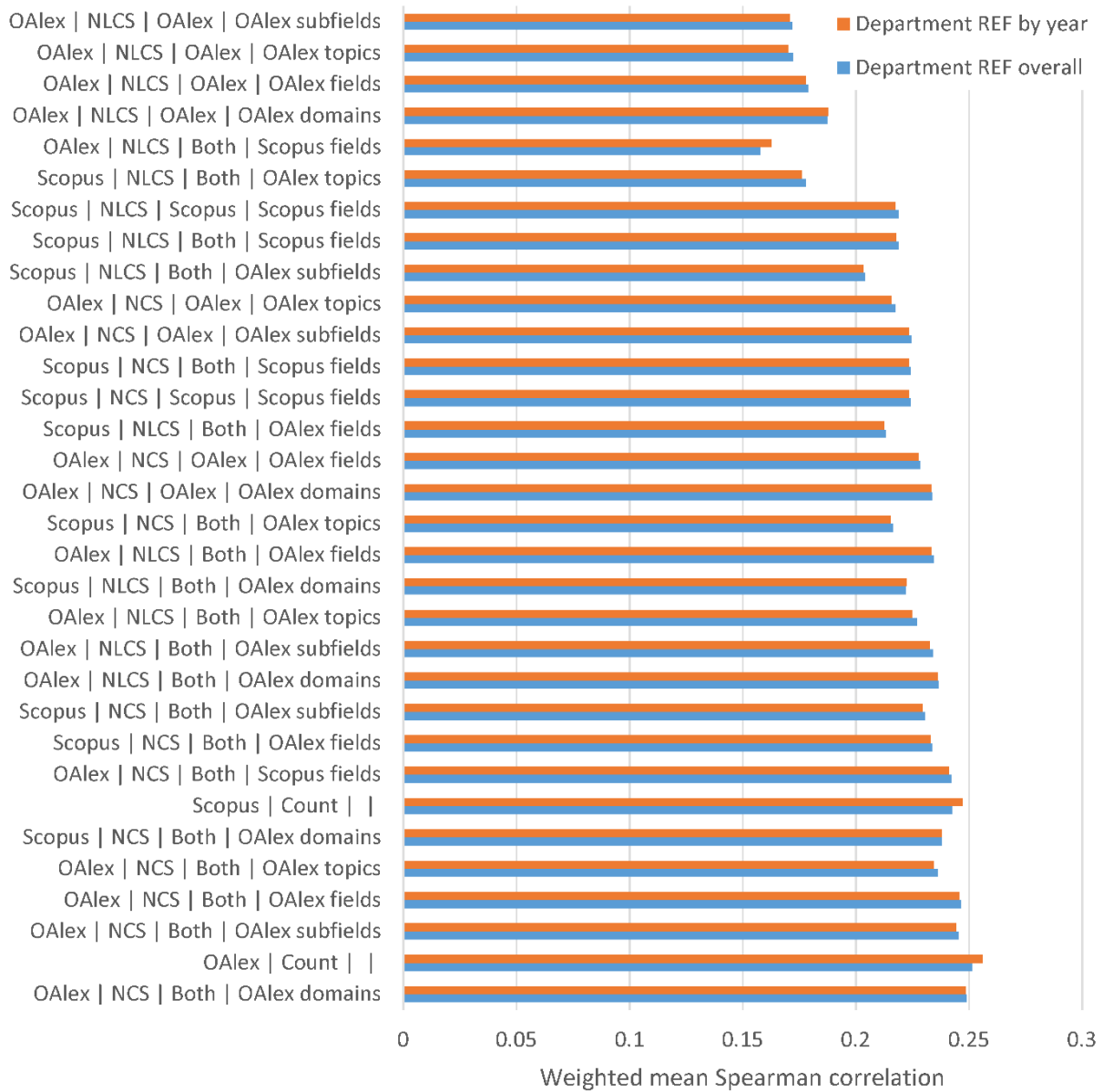

Figure 2. Weighted mean Spearman correlations between indicator values and **departmental average scores** for all qualifying REF articles, where the mean is taken over all 34 UoA correlations, weighted by the number of articles assessed in the UoA. The y-axis order is the same as for Figure 1. The "overall" variant is based on all years combined within each UoA (i.e., one correlation per UoA), and the "by year" variant involves the correlations calculated separately for each UoA and year (i.e., one correlation per year x UoA), and averaged.

If all UoAs are combined for a single correlation with the main ChatGPT gold standard, rather than averaging the correlations calculated separately for each UoA, then the results change substantially (Figure 3). Surprisingly, simple counts are the best indicator overall (i.e., with the highest correlation with ChatGPT scores), despite the disadvantage that they give to more recently published articles. Some correlations for NLCS applied to OpenAlex data are very weak.

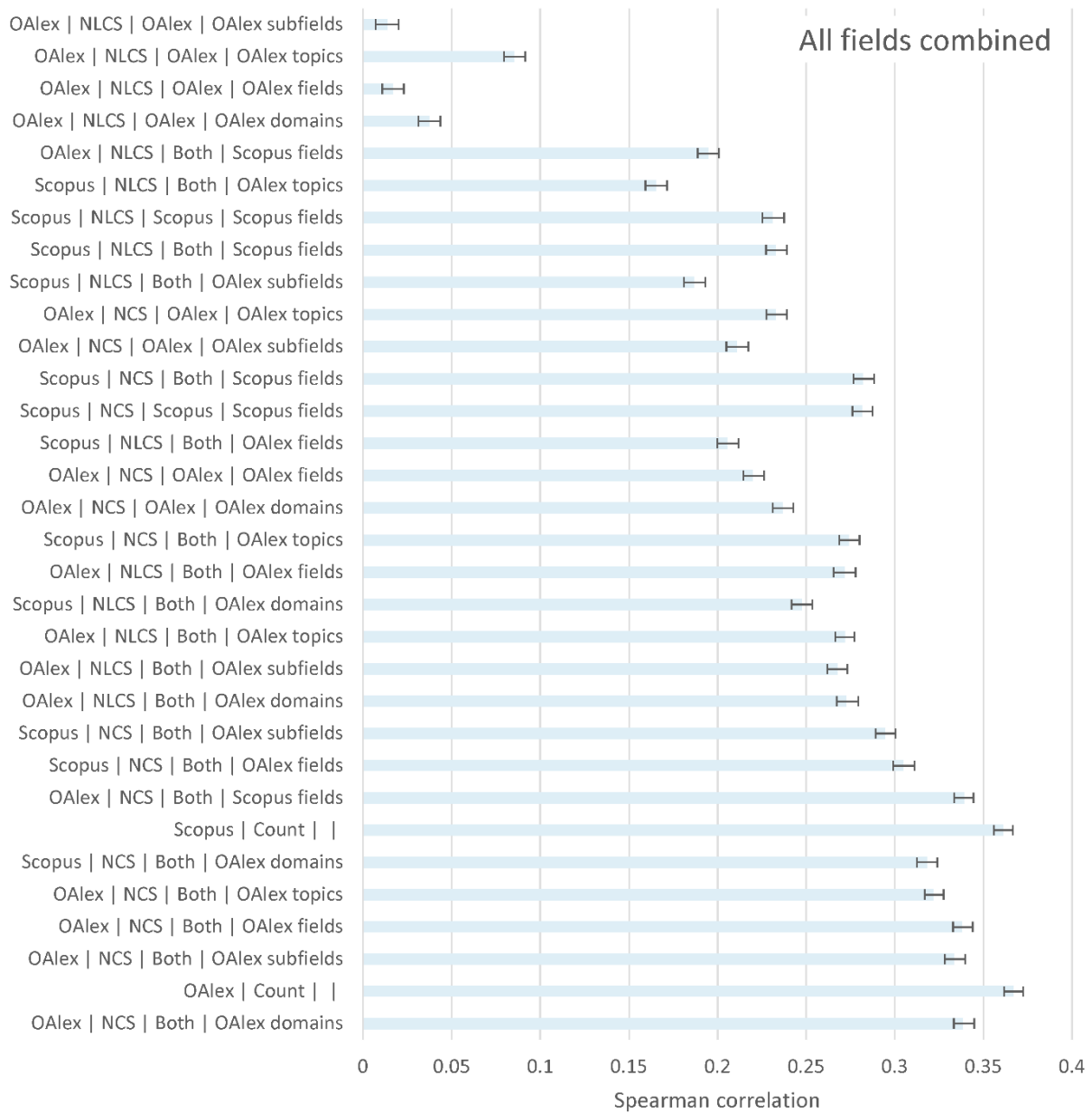

Figure 3. Spearman correlations between indicator values and ChatGPT scores for all qualifying REF articles. Error bars indicate 95% confidence intervals. The y-axis order is the same as for Figure 1. Note the different x axis scale to the previous figures.

*Differences between UoAs in relative indicator values (RQ6)*

The patterns in the correlations with ChatGPT scores by indicator for individual UoAs (see below) differ substantially from the overall pattern for the correlations when all years are combined (Figure 1, blue bars, which only fits the four UoAs of pattern 7 in Table 4; see below). Comparing the overall graph shapes is difficult because their confidence intervals are wide, but the graphs can be clustered heuristically into nine approximate patterns to illustrate some differences (Table 4). In these descriptions, "increasing" means a general tendency for correlations to be higher when they are lower down in the y axis of Figure 1.

Table 4. Approximate patterns in graphs of the correlations between the 32 indicators and the ChatGPT gold standard for the 34 UoAs.

| ID | Pattern | UoAs following this pattern |
|----|---------|------------------------------|
| 1 | All correlations have similar strengths, with mostly overlapping 95% confidence intervals | 6 UoAs: Chemistry; Physics; Mathematical Sciences; Engineering; Politics and International Studies; Area Studies (Figure 4) |
| 2 | As pattern 1 but the first 2 correlations (i.e., the 2 topmost bars) weak. | 1: Law (Figure 5) |
| 3 | As pattern 1 but the first 3 correlations weak. | 1: Business and Management Studies (Figure 5) |
| 4 | As pattern 1 but the first 4 correlations weak. | 2: Architecture, Built Environment and Planning; Economics and Econometrics (Figure 6) |
| 5 | As pattern 1 but the first 5 correlations weak. | 4: Public Health, Health Services and Primary Care; History; Education; Art and Design: History, Practice and Theory (Figure 7) |
| 6 | First 4 correlations weak, then generally increasing. | 1: Geography and Environmental Studies (Figure 8) |
| 7 | First 5 correlations weak, then generally increasing. | 4: Sport and Exercise Sciences, Leisure and Tourism; Clinical Medicine; Allied Health Professions, Dentistry, Nursing and Pharmacy; Psychology, Psychiatry and Neuroscience (Figure 9) |
| 8 | Generally increasing, although at rates varying between UoAs | 11: Biological Sciences; Agriculture, Food and Veterinary Sciences; Earth Systems and Environmental Sciences; Archaeology; Social Work and Social Policy; Sociology; Anthropology and Development Studies; Modern Languages and Linguistics; English Language and Literature; Music, Drama, Dance, Performing Arts, Film and Screen Studies; Communication, Cultural and Media Studies, Library and Information Management (Figure 11) |
| 9 | Highly variable | 4: Computer Science and Informatics; Classics; Philosophy; Theology and Religious Studies (Figure 12) |

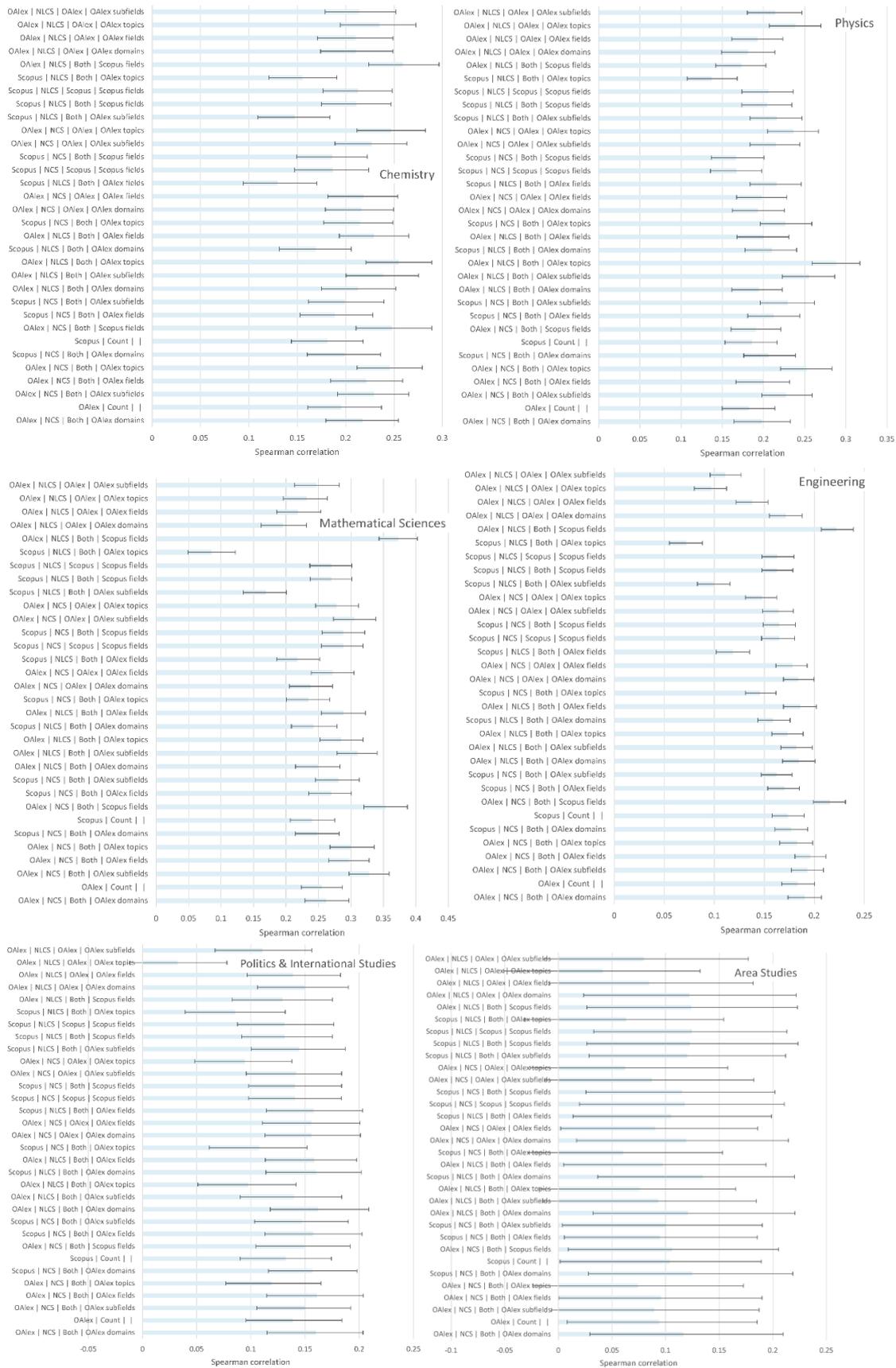

Figure 4. Spearman correlations between indicator values and ChatGPT scores for UoAs matching Table 1, pattern 1: All correlations have similar strengths, with mostly overlapping

95% confidence intervals. Error bars indicate 95% confidence intervals. The y-axis order is the same as for Figure 1. Note the differing x axis scales.

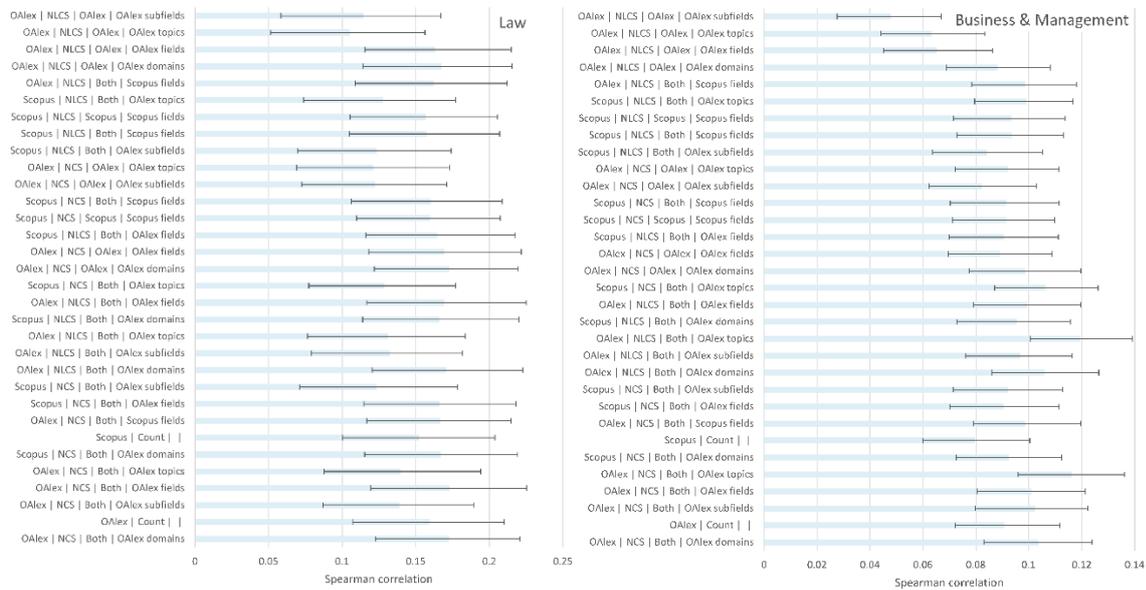

Figure 5. Spearman correlations between indicator values and ChatGPT scores for UoAs matching Table 1, pattern 2 (left) and pattern 3 (right): All correlations have similar strengths, with mostly overlapping 95% confidence intervals, except that the first two/three correlations are weak (topmost bars). Error bars indicate 95% confidence intervals.

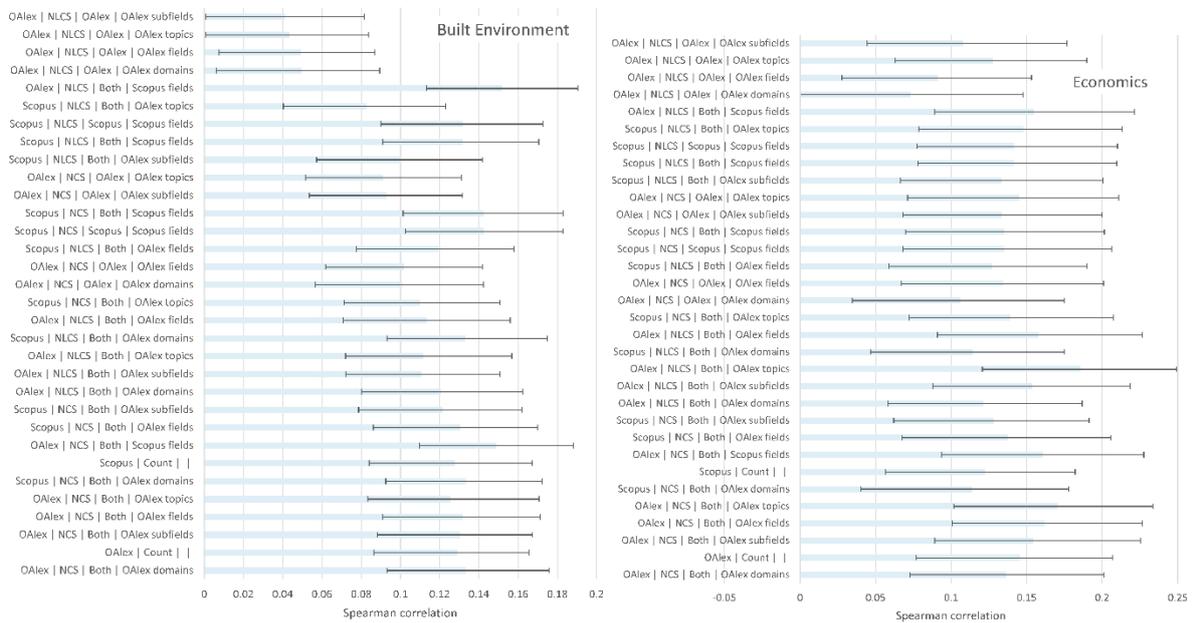

Figure 6. Spearman correlations between indicator values and ChatGPT scores for UoAs matching Table 1, pattern 4: All correlations have similar strengths, with mostly overlapping 95% confidence intervals, except that the first four correlations are weak (topmost four bars). Error bars indicate 95% confidence intervals. The y-axis order is the same as for Figure 1. Note the differing x axis scales.

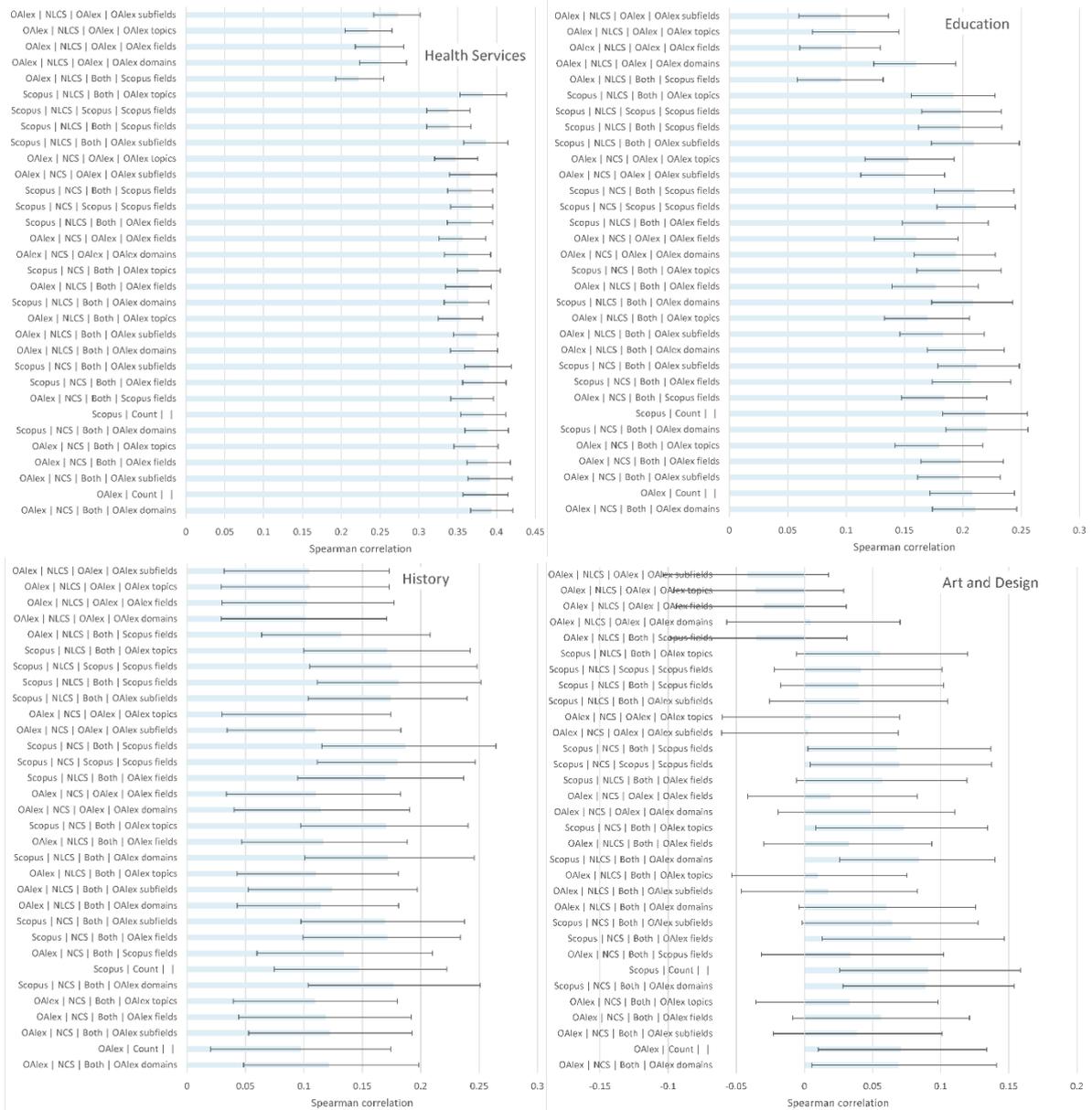

Figure 7. Spearman correlations between indicator values and ChatGPT scores for UoAs matching Table 1, pattern 5: All correlations have similar strengths, with mostly overlapping 95% confidence intervals, except that the first five correlations are weak (topmost bars). Error bars indicate 95% confidence intervals. The y-axis order is the same as for Figure 1. Note the differing x axis scales.

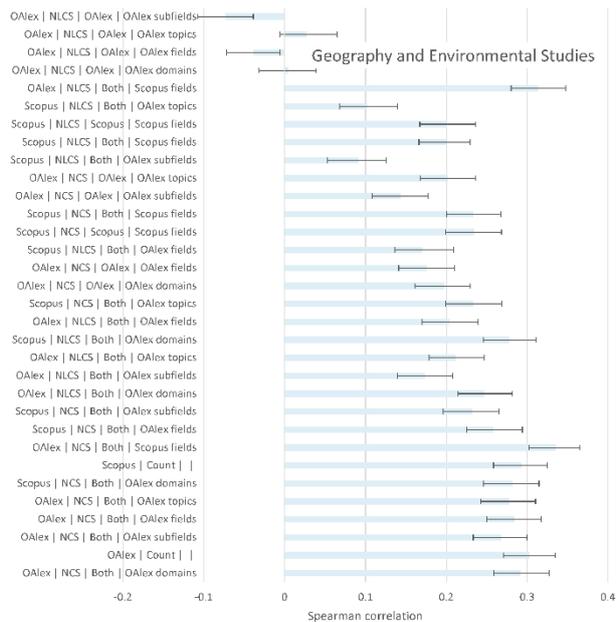

Figure 8. Spearman correlations between indicator values and ChatGPT scores for UoAs matching Table 1, pattern 6: First 4 correlations weak, then generally increasing. Error bars indicate 95% confidence intervals. The y-axis order is the same as for Figure 1. Note the differing x axis scales.

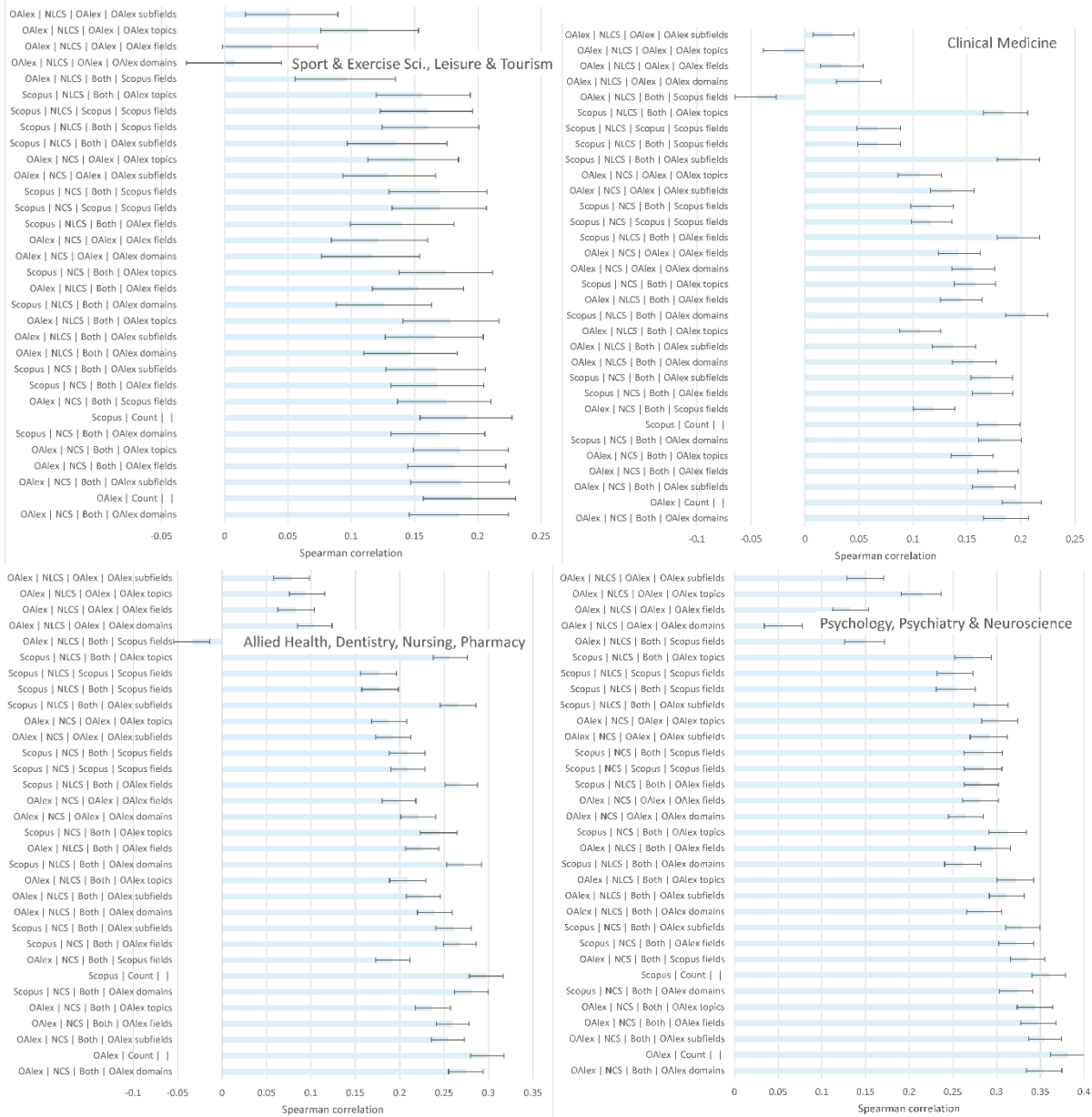

Figure 9. Spearman correlations between indicator values and ChatGPT scores for UoAs matching Table 1, pattern 7: First 5 correlations weak, then generally increasing. Error bars indicate 95% confidence intervals. The y-axis order is the same as for Figure 1. Note the differing x axis scales.

Figure: Spearman correlations across indicator variants for six fields (Sociology, English Lang. & Lit., Social Work, Earth Sys. & Env. Sci, Communication, Biological Sci.).

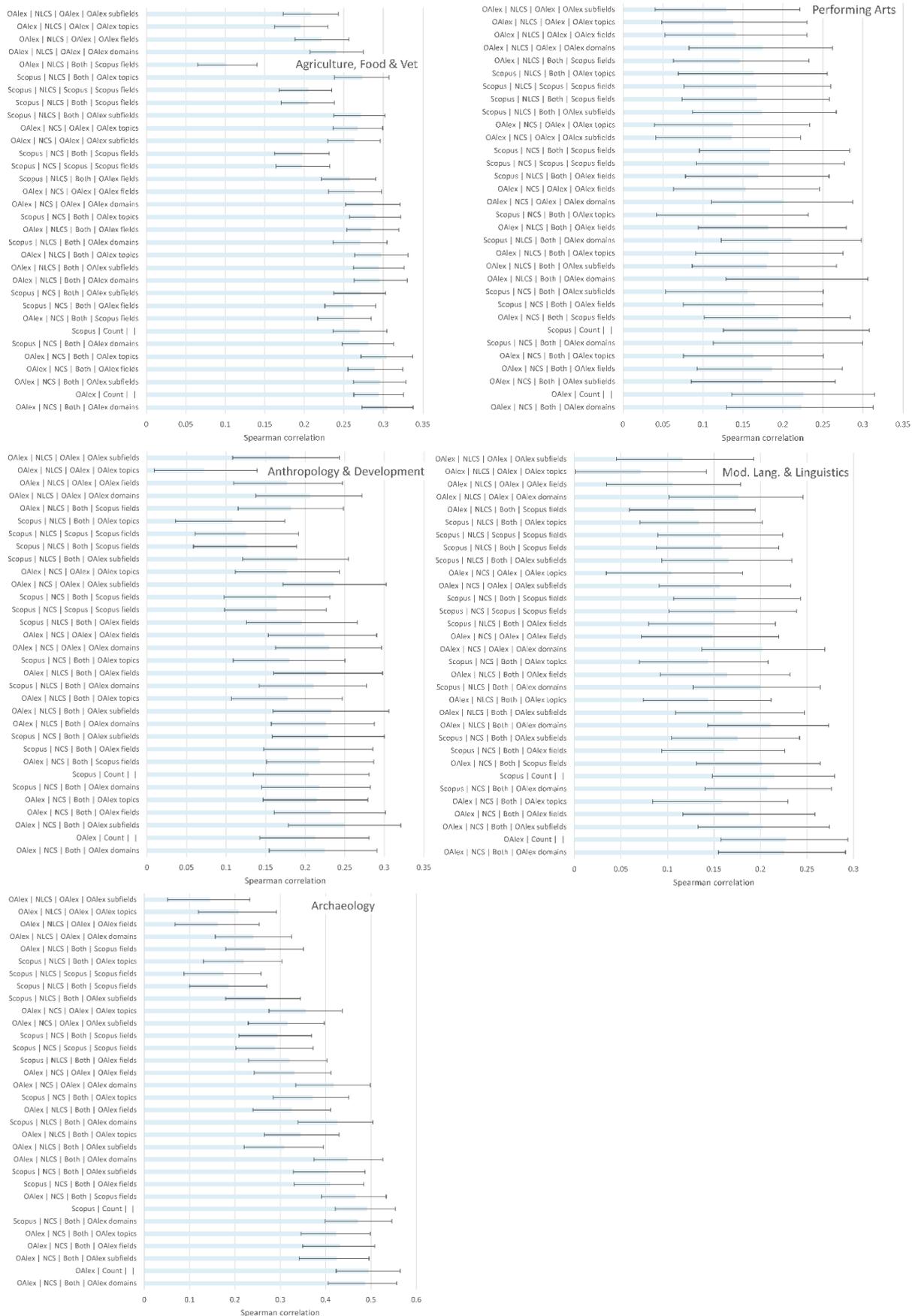

Figure 10. Spearman correlations between indicator values and ChatGPT scores for UoAs matching Table 1, pattern 8: Generally increasing, although at rates varying between UoAs.

Error bars indicate 95% confidence intervals. The y-axis order is the same as for Figure 1. Note the differing x axis scales.

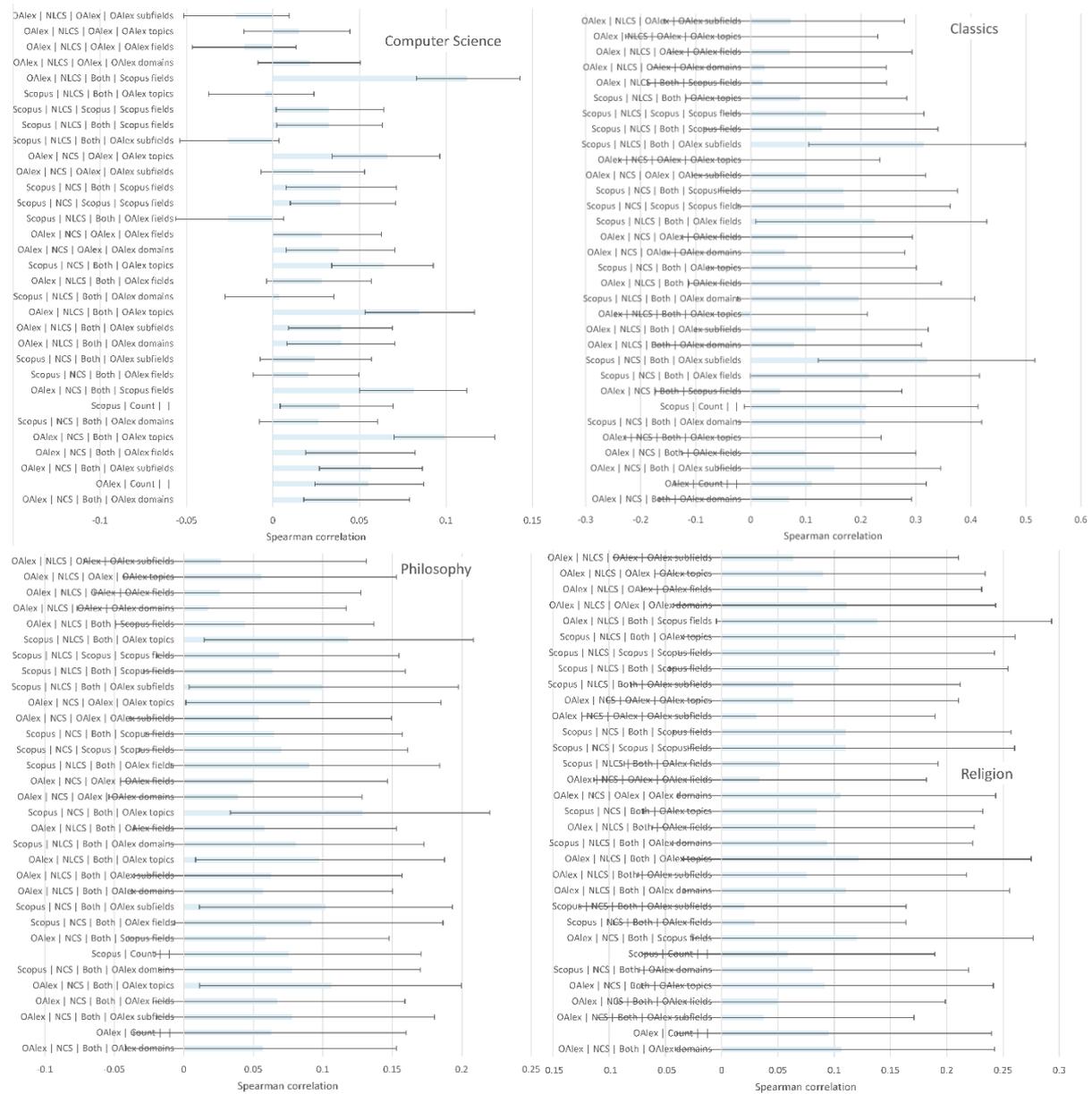

Figure 11. Spearman correlations between indicator values and ChatGPT scores for UoAs matching Table 1, pattern 9: Highly variable. Error bars indicate 95% confidence intervals. The y-axis order is the same as for Figure 1. Note the differing x axis scales.

## Discussion

### Limitations

The results are limited to the particular download of OpenAlex and it seems likely that its contents, indexing technology and classifications will evolve over time. The restriction to articles with DOIs is also a drawback, and OpenAlex might index many low-quality documents without DOIs. Indexing additional low-quality documents unequally between fields would give a field normalised indicator advantage to articles in fields where they were prevalent because they would reduce the mean citation rate for those fields, increasing the normalised indicator values for genuine articles by comparison.

The ChatGPT scores have unknown biases, and, most importantly, may be influenced by citation counts. REF scores are also relatively coarse grained and the use of departmental average score proxies has a damping effect on the magnitude of correlations for the secondary gold standard. The ChatGPT quality scores also have field biases relative to the REF, tending to give higher scores to some UoAs than others, with the differences seeming to favour more highly cited fields (Thelwall & Kurt, 2024). Thus, whilst the known between-UoA biases do not directly influence the within-UoA correlations reported here, their existence suggests that there may well also be within-UoA citation biases in the ChatGPT scores, although their extent is unknown. It is therefore unsafe to draw strong conclusions about the relative strengths of citation counts, NCS and NLCS. Moreover, the citation counts used here are relatively mature, with almost all being at least five years old, and the influence of age would be sharper for newer articles. Using newer data would therefore tend to reduce the citation count correlation and possibly the NCS correlation relative to the others.

## *Comparison with prior research*

The results are broadly consistent with two small-scale prior studies in the sense of the relatively high correlation for raw citation counts compared to other indicators for multi-year data (Bornmann & Leydesdorff, 2013; Fazel & Wolf, 2017) but cannot be directly compared to the other studies cited in the introduction. The results also align with the superiority of article-based subject classification over journal-based subject classification (Klavans & Boyack, 2017).

## *Answers to research questions*

### RQ1: Is OpenAlex better than Scopus for citation counts?

The answer to this is yes because indicators based on its data have the highest correlations with the gold standards, albeit marginally and with the statistical advantage of more indicators in the 32 being compared (so more chances to take advantage of random factors in the data). Since Scopus is more mature, it seems possible that its citations have been more accurately extracted. Its citation sources are presumably more consistently high quality. Nevertheless, these hypothesised advantages have not compensated for the higher citation counts overall for OpenAlex (OpenAlex citation counts were higher for 75% of the gold standard articles and lower for 15%), obtained from its greater coverage.

### RQ2: Which is the best research quality formula: counts, NCS or NLCS?

The results (Figure 1, orange bars, representing seven years of articles analysed together) suggest that the best individual indicator for UoAs (and perhaps therefore broad fields or subject categories) is NCS, closely followed by citation counts, and with NLCS being the worst. Taking into account the fact that many variants of NCS were calculated and that citation counts almost equalled the best NCS, it seems reasonable to consider this to be effectively a tie between citation counts and NCS to offset the multiple chances for NCS. Moreover, for science as a whole with all UoAs merged into a single set for analysis, raw citation counts are best (Figure 3), and when years are analysed separately, raw citation counts are also the best (Figure 1, blue bars, representing articles only being compared against others from the same year). This is surprising because the advantage of NLCS and NCS seem intuitively to be clear and substantial, but the data suggests, at least for the classification schemes and databases used here, that the only advantage of field normalisation formulae is that they also normalise for publication year.

### RQ3: Which is the best subject classification scheme?

The best subject classification scheme was the OpenAlex domains, with the order being mostly Domains>fields>subfields>topics>Scopus Fields. Thus, there is a tendency for the article-based OpenAlex subject classifications to work better than the Scopus journal-based multiple classifications, and, within OpenAlex, for a small number (4) of large categories to be optimal (i.e., its domains). At this level of granularity and taking into account the correlation similarity with raw citation counts, it seems that field normalisation, as traditionally explained and justified implicitly on a much finer grained level (e.g., Glanzel et al., 2009; Waltman et al., 2011), is not working. This issue is discussed again below.

### RQ4: Does OpenAlex or Scopus have the best coverage of journal articles for field normalisation formulae?

This question can't be directly answered. The optimal strategy was to only include documents classified by both sources as journal articles. It isn't possible to directly compare the OpenAlex and Scopus document type classifications because not all other factors can be held constant when comparing these. Nevertheless, whilst OpenAlex seems to perform better than Scopus on most parameters, excluding articles not indexed in Scopus provides a substantial improvement.

### RQ5: Which is the best overall approach for calculating research quality indicators from OpenAlex or Scopus?

The overall best approach for UoAs over the seven year period was to use citation counts from OpenAlex with the NCS formula, classifying documents by OpenAlex domain, but restricting the scope of the documents to those that are classified as journal articles by Scopus and OpenAlex. Raw citation counts from OpenAlex were best for all UoAs combined, and for individual UoAs and years.

### RQ6: Are there field differences in the answer to RQ5?

There are substantial differences between UoAs in the overall trends, as evidenced by the nine differing patterns identified. For example, NLCS is the best formula for Clinical Medicine and the narrowest classification, OpenAlex topics, is best for Physics, at least in terms of correlations with ChatGPT scores.

### *Is OpenAlex suitable for citation-based indicators?*

Although the optimal citation-based indicator combines information from both Scopus and OpenAlex, the second best one, OpenAlex citation counts (Figure 1) are almost as good, so OpenAlex is technically suitable for constructing citation-based indicators. If only field normalised indicators are considered, then the best OpenAlex-only indicator (OAlex|NCS|OAlex |OAlex domains) is substantially weaker but still reasonably strong in terms of correlation with average ChatGPT scores.

## Is field normalisation needed?

Surprisingly, the results suggest that field and year normalisation is unnecessary for articles with mature citation counts. Although the strongest overall broad field (i.e., UoAs) correlation used NCS rather than raw citation counts, the OpenAlex domains used encompass only four classes and the difference in correlations is only marginal, and was only evident for the seven years combined, rather than for the years treated separately. Thus, year normalisation is necessary but not field normalisation.

It seems reasonable to expect that normalisation would still be necessary for sets of articles where there was a range of ages that included relatively new articles, such as less than three years old, as well as older articles. It also seems reasonable to speculate that, in general, for articles with mature citation counts, expert quality judgements correlate better with raw citation counts than with normalised citation counts based on appropriately narrow (e.g., 27+ fields) fields.

The strength of raw citation counts as a research quality indicator that is almost as effective as the best field normalised citation formula tested is surprising given that field normalisation is a central tenet of scientometrics. Its value seems to be driven by the belief that low citation specialties can be as important as high citation specialties because their lower counts reflect factors irrelevant to research quality, such as short reference lists, a tendency to cite non-article documents including books, or an applied orientation with non-scholarly impacts. The value of field normalisation never seems to have been compared with raw citation counts against an independent indicator of research quality at scale before, so its relative weakness is a new finding. There are many possible explanations for this result, including the following.

- All subject classification schemes tested are ineffective at grouping together articles from similar specialties, so whilst field normalisation is a good idea in theory, in practice it does not work with existing subject classification schemes – or at least the ones tested here from Scopus and OpenAlex.
- There is a strong tendency for more cited specialties to produce research that experts tend to regard as higher quality. At the journal level, for example, this would allow Journal Impact Factors (JIFs) to be compared between fields and all of the most cited journals in a low citation field might tend to publish lower quality research than all least cited journals in a high citation field.

Both above may be true to some extent. It seems reasonable to consider the possibility that a more accurate subject classification scheme might produce better field normalised indicators, and this is a challenge for future research. Existing classification schemes may also work better, such as that of Dimensions or the Web of Science, as might a different way of applying the Scopus or OpenAlex schemes. Nevertheless, it also seems likely, based on the results, that field normalisation at a reasonable level of granularity (e.g., 27+ fields) goes too far in equalising between specialties when there are genuine average quality differences between them. One prior study has suggested that there may be field differences in quality standards, at least for interdisciplinary research (Thelwall et al., 2023b). To give an extreme example, it seems plausible that almost all clinical medicine would be regarded by most academics as better than all scientometrics because it tends to be more expensive, larger scale, with greater ethical concerns, and a bigger (at least direct) influence on humanity. Field normalising citations to allow the two to be compared, whether with NCS or NLCS, may go too far in equalising them for research evaluations.

### Broad or narrow subject categories for NLCS and NCS?

The results suggest that broader subject categories are better for field normalisation, perhaps because narrow categories go too far in equalising between specialties.

### NLCS or NCS?

NCS seems to be better than NLCS, although not for all fields. This is surprising because NLCS was designed to solve the statistical problem with NCS that its values could be unduly influenced by a minority of highly cited articles in the denominator of the calculations. A possible explanation is that a side-effect of the NLCS reducing the importance of highly cited articles to the denominator of the NLCS formula is that it increases the relative importance of uncited articles. This can cause problems if there are many uncited articles that are not genuine articles but are other uncitable document types, such as editorials, or less cited document types, like short form articles.

NCS seems to be better than NLCS in both Scopus and OpenAlex, so document classification issues may affect each one. It is not clear whether this problem is due to errors and thus avoidable in theory. Whilst there are clear problems with OpenAlex's coarse document type classification scheme (i.e., everything in a journal is an article), but not necessarily in Scopus, although it seems to index too many uncited articles (Thelwall, 2016). This could also be a more fundamental issue of uncitable or rarely cited documents occurring in the scientific record that are nevertheless valid journal articles. Another possibility is that mixed article lengths cause a much higher proportion of uncited articles than would be the case if all had similar lengths.

## Conclusions

Although the results are not conclusive due to a reliance on ChatGPT with unknown biases and indirect and age-biased expert quality scores from the REF, they suggest that OpenAlex is suitable for citation analysis, especially if its document type classifications are helped by Scopus. The graphs for the different UoAs above may help to suggest which variant of NLCS or NCS and which classification scheme would be most suitable.

The unexpected result that field normalisation has little or no value for mature citation data, at least in the configurations tested here (Scopus multiple field classifications, OpenAlex primary classifications), is counterintuitive and needs further exploration before changes are made to practice. Nevertheless, it should at least raise the possibility that one of the effects of field normalisation is to equalise between fields or specialties that are genuinely different in average research quality. Thus, scientometricians should not assume that the most cited from one field are as good as the most cited from another and should accept the possibility that there are genuine differences in average quality levels between fields and between specialties. Given that quality standards and criteria differ between fields, making direct comparisons to verify this would be difficult. Whilst it might be rare to directly compare very different fields in a single research evaluation, this can occur when multidisciplinary departments are assessed or when a department includes supporting specialties, such as statistics.

Despite the above conclusions, for policy reasons, research evaluators might take the perspective that it is necessary to start with the assumption that all fields or specialties are equal and design their systems to ensure that the results reflect this. For example, in the REF, it would cause substantial arguments if the average scores were allowed to differ substantially between UoAs. On this basis, field normalisation might be used as a tool to help achieve an imaginary equality between fields/specialties rather than to reflect an existing equality.

# References

Alperin, J. P., Portenoy, J., Demes, K., Larivière, V., & Haustein, S. (2024). An analysis of the suitability of OpenAlex for bibliometric analyses. arXiv. https://doi.org/10.48550/arXiv.2404.17663

Bornmann, L., & Leydesdorff, L. (2013). The validation of (advanced) bibliometric indicators through peer assessments: A comparative study using data from InCites and F1000. Journal of Informetrics, 7(2), 286-291.

Bornmann, L., & Marx, W. (2015). Methods for the generation of normalized citation impact scores in bibliometrics: Which method best reflects the judgements of experts? Journal of Informetrics, 9(2), 408-418.

Brzezinski, M. (2015). Power laws in citation distributions: evidence from Scopus. Scientometrics, 103, 213-228.

Céspedes, L., Kozlowski, D., Pradier, C., Sainte‑Marie, M. H., Shokida, N. S., Benz, P., & Larivière, V. (2025). Evaluating the linguistic coverage of OpenAlex: An assessment of metadata accuracy and completeness. Journal of the Association for Information Science and Technology. https://doi.org/10.1002/asi.24979

Culbert, J., Hobert, A., Jahn, N., Haupka, N., Schmidt, M., Donner, P., & Mayr, P. (2024). Reference coverage analysis of OpenAlex compared to Web of Science and Scopus. arXiv. https://doi.org/10.48550/arXiv.2401.16359

De Bellis, N. (2009). *Bibliometrics and citation analysis: from the science citation index to cybermetrics*. Scarecrow press.

Fazel, S., & Wolf, A. (2017). What is the impact of a research publication? BMJ Mental Health, 20(2), 33-34.

Glänzel, W., Thijs, B., Schubert, A., & Debackere, K. (2009). Subfield-specific normalized relative indicators and a new generation of relational charts: Methodological foundations illustrated on the assessment of institutional research performance. Scientometrics, 78, 165-188.

Haupka, N., Culbert, J. H., Schniedermann, A., Jahn, N., & Mayr, P. (2024). Analysis of the publication and document types in OpenAlex, Web of Science, Scopus, PubMed and Semantic Scholar. *arXiv preprint arXiv:2406.15154*.

Jiao, C., Li, K., & Fang, Z. (2023). How are exclusively data journals indexed in major scholarly databases? An examination of the Web of Science, Scopus, Dimensions, and OpenAlex. *arXiv preprint arXiv:2307.09704*.

Klavans, R., & Boyack, K. W. (2017). Which type of citation analysis generates the most accurate taxonomy of scientific and technical knowledge? Journal of the Association for Information Science and Technology, 68(4), 984-998.

Langfeldt, L., Nedeva, M., Sörlin, S., & Thomas, D. A. (2020). Co-existing notions of research quality: A framework to study context-specific understandings of good research. Minerva, 58(1), 115-137.

Lundberg, J. (2007). Lifting the crown - citation z-score. Journal of informetrics, 1(2), 145-154.

Maddi, A., Maisonobe, M., & Boukacem-Zeghmouri, C. (2024). Geographical and Disciplinary Coverage of Open Access Journals: OpenAlex, Scopus and WoS. *arXiv preprint arXiv:2411.03325*.

Moed, H. F. (2006). Citation analysis in research evaluation. Springer Science & Business Media.

Milyaeva, S., & Neyland, D. (2023). Let's agree to agree: The situational academic quality of the UK REF as consensual public knowledge. Social Studies of Science, 53(3), 427-448.

OpenAlex (2025). Topics. https://docs.openalex.org/api-entities/topics

Priem, J., Piwowar, H., & Orr, R. (2022). OpenAlex: A fully-open index of scholarly works, authors, venues, institutions, and concepts. arXiv preprint arXiv:2205.01833.

Purkayastha, A., Palmaro, E., Falk-Krzesinski, H. J., & Baas, J. (2019). Comparison of two article-level, field-independent citation metrics: Field-Weighted Citation Impact (FWCI) and Relative Citation Ratio (RCR). Journal of Informetrics, 13(2), 635-642.

Thelwall, M. (2016). Are there too many uncited articles? Zero inflated variants of the discretised lognormal and hooked power law distributions. Journal of Informetrics, 10(2), 622-633. https://doi.org/10.1016/j.joi.2016.04.014

Thelwall, M. (2017). Three practical field normalised alternative indicator formulae for research evaluation. *Journal of Informetrics*, *11*(1), 128-151.

Thelwall, M. (2024). Can ChatGPT evaluate research quality? Journal of Data and Information Science, 9(2), 1–21. https://doi.org/10.2478/jdis-2024-0013

Thelwall, M. (2025). Evaluating research quality with Large Language Models: An analysis of ChatGPT's effectiveness with different settings and inputs. Journal of Data and Information Science, 10(1), 1-19. https://doi.org/10.2478/jdis-2025-0011

Thelwall, M., Jiang, X., & Bath, P. A. (2024). Evaluating the quality of published medical research with ChatGPT. *arXiv preprint arXiv:2411.01952*.

Thelwall, M., Kousha, K., Stuart, E., Makita, M., Abdoli, M., Wilson, P., & Levitt, J. (2023a). In which fields are citations indicators of research quality? *Journal of the Association for Information Science and Technology*, *74*(8), 941-953.

Thelwall, M., Kousha, K., Stuart, E., Makita, M., Abdoli, M., Wilson, P. & Levitt, J. (2023b). Does the perceived quality of interdisciplinary research vary between fields? Journal of Documentation. 79(6), 1514-1531. https://doi.org/10.1108/JD-01-2023-0012

Thelwall, M., & Kurt, Z. (2024). Research evaluation with ChatGPT: Is it age, country, length, or field biased? arXiv preprint arXiv:2411.09768.

Thelwall, M., & Yaghi, A. (2024). In which fields can ChatGPT detect journal article quality? An evaluation of REF2021 results. *arXiv preprint arXiv:2409.16695*.

van Raan, A. F. (2004). Measuring science. Handbook of quantitative science and technology research. Springer (pp. 19-50).

Waltman, L., van Eck, N. J., van Leeuwen, T. N., Visser, M. S., & van Raan, A. F. (2011). Towards a new crown indicator: Some theoretical considerations. Journal of Informetrics, 5(1), 37-47.

Waltman, L., Calero‐Medina, C., Kosten, J., Noyons, E. C., Tijssen, R. J., van Eck, N. J., & Wouters, P. (2012). The Leiden Ranking 2011/2012: Data collection, indicators, and interpretation. Journal of the American Society for Information Science And Technology, 63(12), 2419-2432.

Wilsdon, J., Allen, L., Belfiore, E., Campbell, P., Curry, S., Hill, S., & Viney, I. (2015). The metric tide: report of the independent review of the role of metrics in research assessment and management. https://www.ukri.org/wp-content/uploads/2021/12/RE-151221-TheMetricTideFullReport2015.pdf